\documentclass[]{jfm}%lineno

\usepackage{graphicx}
\usepackage{graphbox}
\usepackage{newtxtext}
\usepackage{newtxmath}
\usepackage{mathtools}

\usepackage{natbib}
\usepackage{hyperref}
\hypersetup{
    colorlinks = true,
    urlcolor   = blue,
    citecolor  = black,
}

\usepackage{multirow}
\usepackage{accents}
\usepackage{comment}
\usepackage{subcaption}
\usepackage{xcolor} 
\usepackage{epstopdf, epsfig}
\usepackage{tikz}

\newcommand{\RomanNumeralCaps}[1]
\linenumbers
\graphicspath{{figures/}}
\usepackage{cleveref}

\shorttitle{PINN to augment experimental data}

\shortauthor{Zhu, Jiang, Lefauve, Kerswell, and Linden}

\title{Physics-informed neural network to augment experimental data: an application to stratified flows}

\author{Lu Zhu\aff{1} \corresp{\email{lz447@cam.ac.uk}}, Xianyang Jiang\aff{1}, 
     Adrien Lefauve\aff{1}, 
     Rich R. Kerswell\aff{1}, P. F. Linden\aff{1}}

 \affiliation{\aff{1}Department of Applied Mathematics and Theoretical Physics, University of Cambridge, Wilberforce Road, Cambridge CB3 0WA, UK}

\begin{document}

\maketitle

\begin{abstract}
We develop a physics-informed neural network (PINN) to significantly augment state-of-the-art experimental data and apply it to stratified flows. The PINN is a fully-connected deep neural network fed with time-resolved, three-component velocity fields and density fields measured simultaneously in three dimensions at $Re = O(10^3)$ in a stratified inclined duct experiment. The PINN enforces incompressibility, the governing equations for momentum and buoyancy, and the boundary conditions by automatic differentiation. The physics-constrained, augmented data are output at an increased spatio-temporal resolution and demonstrate five key results: (i) the elimination of measurement noise; (ii) the correction of distortion caused by the scanning measurement technique; (iii) the identification of weak but dynamically important three-dimensional vortices; (iv) the revision of turbulent energy budgets and mixing efficiency; and (v) the prediction of the latent pressure field and its role in the observed Holmboe wave dynamics. These results mark a significant step forward in furthering the reach of experiments, especially in the context of turbulence, where accurately computing three-dimensional gradients and resolving small scales remain enduring challenges.

\end{abstract}


\section{Introduction}

Since the seminal pipe flow experiments of Osborne Reynolds~\citep{reynolds_experimental_1883}, experiments have been widely designed and used to study fluid flow \citep{tropea2007springer}.
The recent development of state-of-the-art measurement techniques allows the investigation of flow fields at high spatio-temporal resolution (e.g. \citealp{partridge_versatile_2019}). 
However, experimentalists often face the challenge of accurately measuring flow structures across a wide range of scales, particularly in turbulent flows. Moreover, latent flow variables, such as the pressure field, can rarely be measured.
To complement experiments, numerical simulations are widely used, 
often affording better resolution and the full set of flow variables. However, simulations remain idealised models subject to computational limits. These limitations make them challenging to deploy in regions of parameter space appropriate to environmental or industrial applications, as well as in realistic geometries with non-trivial boundary conditions.

Recent advances in machine learning 
have stimulated new efforts in fluid mechanics~\citep{Vinuesa2023nature}, with one application being the reconstruction of flow fields from limited observations \citep{fukami2019super,raissi2020hidden}. Among the available tools, physics-informed neural networks (PINN)~\citep{raissi2019physics} hold particular promise. 
The idea behind a PINN is to impose physical laws on a neural network fed with observations. This allows the model to super-resolve the flow in space and time, to remove spurious  noise, and to predict unmeasured (latent) variables such as the pressure~\citep{raissi2019physics}. 
While PINNs have been used in numerical fluid problems~\citep{cuomo2022scientific}, their application to experiments remains limited, perhaps due to the scarcity of high-quality data.

In this paper, we demonstrate the potential of a PINN to augment experimental data and reveal new physical insights into stratified flows. For this purpose, we use the canonical stratified inclined duct (SID) experiment~\citep{meyer2014stratified} sustaining a salt-stratified shear flow in a long tilted duct.The PINN is fed datasets comprising the time-resolved, three-component velocity field and density field measured simultaneously in a three-dimensional volume. We focus on experiments in the Holmboe wave (HW) regime, as these interfacial waves are important precursors of turbulence in environmental flows, e.g. between salt-stratified layers in the ocean \citep{kawaguchi2022}. 

In the remainder of this paper, we describe the datasets and the PINN in~\S~\ref{sec:method}, and our results in \S~\ref{sec:results}. We demonstrate the improvement in signal-to-noise ratio in \S~\ref{sec:results_quality}, in the detection of weak but important coherent structures in \S~\ref{sec:vortex}, in the accuracy of energy budgets and quantification of mixing in \S~\ref{sec:eng_budget}. We also show  \S~\ref{sec:pres} that the latent pressure field revealed by the PINN is key to explain asymmetric HWs in SID. Finally, we conclude in \S~\ref{sec:ccl}.

\section{Methodology}\label{sec:method}

\subsection{The stratified inclined duct (SID) dataset}\label{sec:method:exp}

\begin{figure}
      \centering    
      \includegraphics[width=0.75\linewidth, trim=0mm 0mm 0mm 0mm, clip]{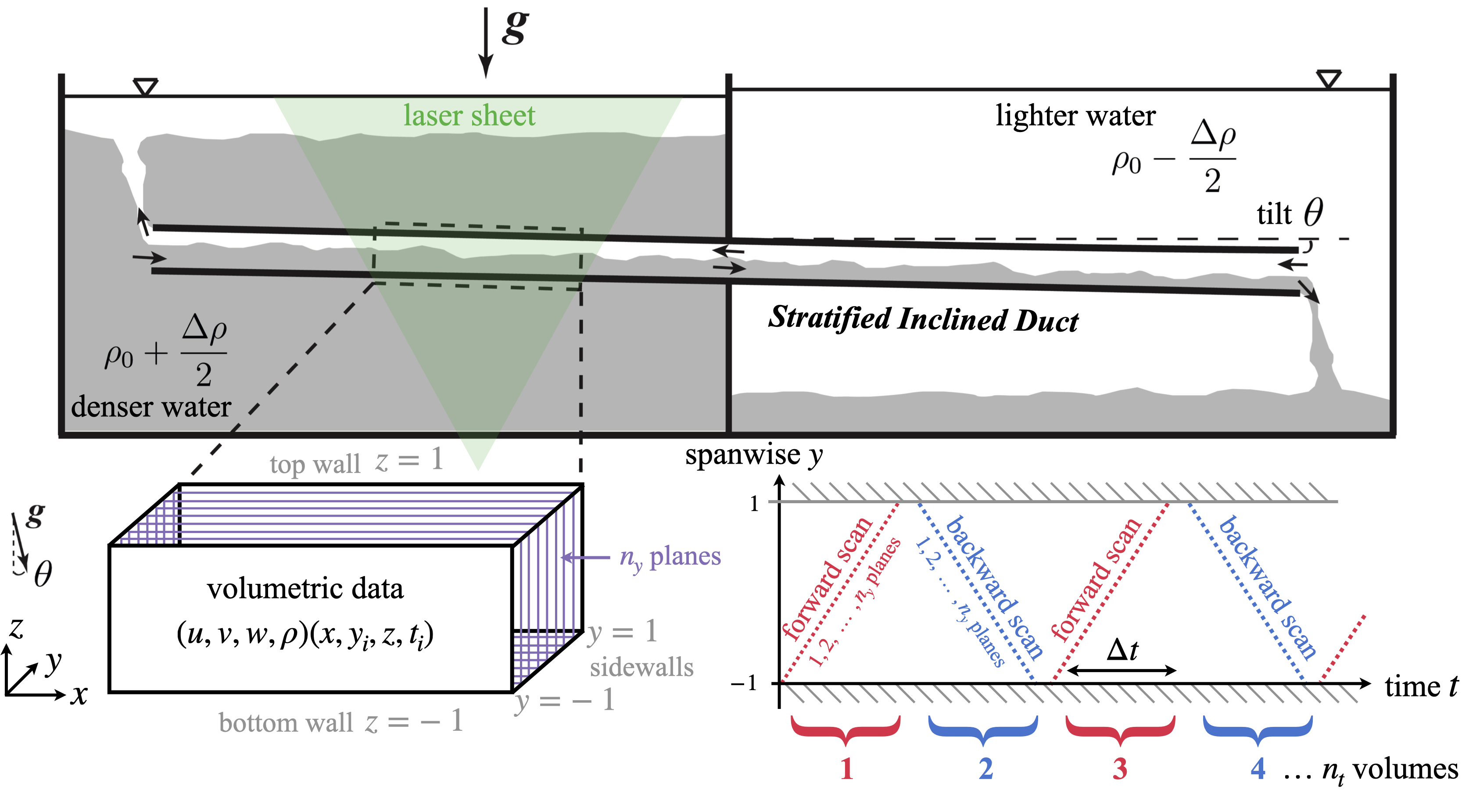}
    \caption{The SID setup and  dataset. Each volume is constructed by aggregating $n_y$ planes (closely-spaced dots) obtained by scanning across the duct over time $\Delta t$}
    \label{fig:sid}
\end{figure}

The data were collected in the stratified inclined duct facility (SID, sketched in figure~\ref{fig:sid}), where two salt solutions with density $\rho_0\pm \Delta\rho/2$ are exchanged through a long square duct tilted at an angle $\theta$. 
Importantly, the volumetric, three-component velocity and density fields
are collected through a continuous back-and-forth scanning of a streamwise ($x$) -- vertical ($z$) laser sheet across the spanwise ($y$) direction, as introduced in \cite{partridge_versatile_2019}. Figure~\ref{fig:sid} shows how $n_y$ successive planar measurements of laser-induced fluorescence (for density) and stereo particle image velocity (for velocity) captured over a short time $\Delta t$ are aggregated to yield $n_t$ `near-instantaneous' volumes. The typical processed dataset has $(n_x,n_y,n_z,n_t) \approx (400,35,80,250)$ points, noting that the spatial resolution in $x,z$ is identical but is about two to three times higher (better) than along $y$. 

All data are made non-dimensional with the following scales. For the spatial coordinates we use the half-duct height $H^*/2=22.5$ mm.  For the velocity we use half the fixed peak-to-peak `buoyancy velocity' scale $U^*/2 \equiv \sqrt{g^\prime H^*}$ (where $g^\prime=g\Delta\rho/\rho_0$ is the reduced gravity chosen for the experiment), leading to velocities being approximately bounded by $\pm 1$. This means time is non-dimensionalised by the advective unit $H^*/U^*$, yielding the Reynolds number $\Rey = H^* U^\ast/4\nu$, where $\nu$ is the kinematic viscosity of water. The Prandtl number is  $\Pran=\nu/\kappa \approx 700$, where $\kappa$ the molecular diffusivity of salt. For the density field (its deviation from the neutral level $\rho_0$, i.e. the buoyancy) we use half the maximum jump $\Delta\rho/2$, which yields the fixed bulk Richardson number $\Ri =  (g^\prime/2) (H^\ast/2)/(U^{\ast}/2)^2=1/4$. The data can be downloaded from ~\cite{lefauve_partridge_linden_2019}.

We focus on comparing two typical Holmboe wave (HW) datasets: H1 featuring a double-mode, symmetric HW at $(\Rey,\theta)=(1455,1^\circ)$; and H4, featuring a single-mode, asymmetric HW at $(\Rey,\theta)=(438,5^\circ)$ studied in detail in \cite{lefauve2018structure}.

\subsection{The physics-informed neural network (PINN)}

\begin{figure}
\centering
    \includegraphics[width=0.8\linewidth, trim=0mm 0mm 0mm 0mm, clip]{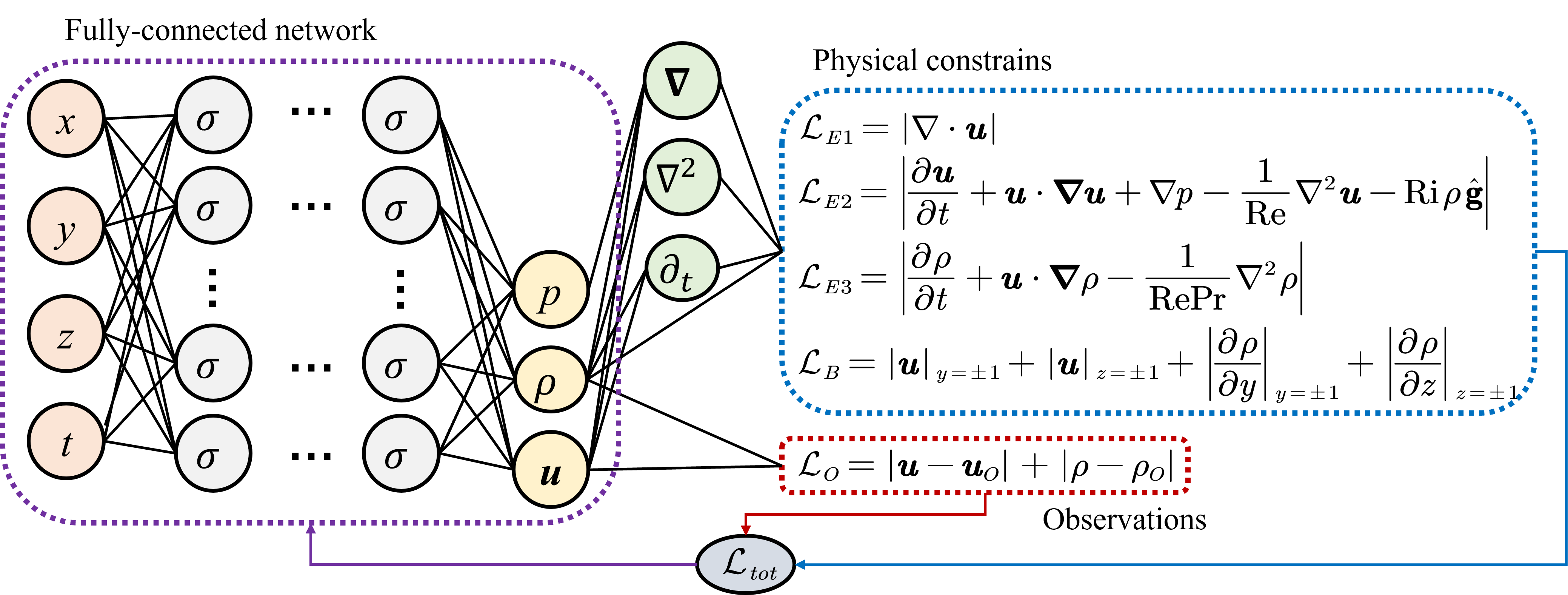}
    \caption{Schematics of the physics-informed neural network (PINN). The output variables $(\boldsymbol{u},\rho,p)$ (in yellow) are predicted from the input variables $(\boldsymbol{x},t)$ (in orange) subject to physical constraints and experimental observations. \vspace{-0.2cm}}
    \label{fig:pinn}
\end{figure}

The PINN is sketched in figure \ref{fig:pinn}. A fully-connected deep neural network is set up using the spatial $\boldsymbol{x}=(x,y,z)$ and temporal $t$ coordinates of the flow domain as the inputs, and the corresponding velocity $\boldsymbol{u}=(u,v,w)$, density $\rho$, and pressure $p$ as the outputs. The network is composed of 14 layers with an increasing number of artificial neurons ([$64\times 4$, $128\times 3$, $256\times 4$, $512\times 3$]) (a sensitivity analysis was conducted to ensure  convergence). The outputs of each layer $\boldsymbol{n}_k$ are computed by a nonlinear transformation of the previous layer $\boldsymbol{n}_{k-1}$ following the basic `neuron' $\boldsymbol{n}_k=\boldsymbol{\sigma}(\boldsymbol{w}^T_{k-1}\boldsymbol{n}_{k-1}+\boldsymbol{b}_{k-1})$, where $\boldsymbol{b}_{k-1}$  and $\boldsymbol{w}^T_{k-1}$ are the bias vectors and weight matrices of layer $k-1$. To introduce nonlinearity and overcome the potential vanishing gradient of deep networks, we use a Swish activation function $\boldsymbol{\sigma}$  for all the hidden layers~\citep{ramachandran2017searching}.

The outputs $\boldsymbol{u}$ and $\rho$ are compared with experimental data $\boldsymbol{u}_O$ and $\rho_O$ to obtain the absolute mean error loss function $\mathcal{L}_O$ (in red in figure~\ref{fig:pinn}). The spatial and temporal derivatives of $\boldsymbol{u}$, $\rho$ and $p$ are computed at every sampling point using automatic differentiation~\citep{baydin2018automatic} (in green).  To impose the physical constraints, the derivatives are substituted into the governing mass, momentum, and density scalar equations,  corresponding to loss functions $\mathcal{L}_{E1}$, $\mathcal{L}_{E2}$, $\mathcal{L}_{E3}$, respectively (in blue). We also impose the boundary conditions through $\mathcal{L}_{B}$ (no-slip $\boldsymbol{u}=0$, and no-flux  $\partial_y\rho=0, \partial_z\rho=0$ at the four walls $y=\pm 1, z=\pm 1$, respectivley).

Combining these constraints, we define the total loss function 
\begin{equation}
\label{eq:loss_func}
\mathcal{L}_\mathrm{tot}=\frac{\lambda_E}{N_E}\sum_{j=0}^{N_E}\sum_{i=0}^3\mathcal{L}_{Ei}^j+
\frac{\lambda_B}{N_B}\sum_{j=0}^{N_B}\mathcal{L}_{B}^j+
\frac{\lambda_O}{N_O}\sum_{j=0}^{N_O}\mathcal{L}_{O}^j,
\end{equation}
where $\lambda_E$, $\lambda_B$, and $\lambda_O$ are weight coefficients for the governing equation, boundary condition and observation losses, respectively. We define the number of training samples $N_E$, $N_B$, and $N_O$ for the equations, boundaries and observations, respectively. Here $N_E=3 \times 10^7$ for H1 and $N_E=2\times 10^7$ for H4 (as it depends on the resolution of the data), $N_B=10^7$ and $N_O=2\times 10^6$. The neural network is trained to seek the optimal parameters using the ADAM algorithm~\citep{kingma2014adam} to minimise the total loss $\mathcal{L}_\mathrm{tot}$. To enhance convergence, we adopt exponentially decaying optimizer steps.

A key strength of this PINN is its natural ability to reconstruct truly instantaneous three-dimensional flow fields by overcoming the spanwise distortion of our `near-instantaneous' data acquired by scanning (as also attempted by \citealp{knutsen2020inter,zigunov2023continuously}). 
This is done simply by
feeding the $n_y$ successive snapshots $(\boldsymbol{u},\rho)(x,y_i,z,t_i)$, taken during each alternating forward and backward scan (see figure~\ref{fig:sid}), at times $t_i$ equally spaced by $\Delta t/n_y$, where $(\Delta t,n_y)=(2.29,39)$ for H1 and $(1.08,30)$ for H4.

\section{Results}\label{sec:results}

\subsection{Improved quality of experimental data}\label{sec:results_quality}

We start by comparing flow snapshots and statistics of HWs in the raw experimental data to the PINN-reconstructed data. Here the spatial resolution of the PINN data is typically doubled in all directions $x,y,z$ compared to the experimental data, which was sufficient for convergence (though higher resolutions are possible by increasing the sampling input points in the PINN). Instantaneous volumes are also generated at a higher temporal resolution than acquired experimentally, namely at intervals $\Delta t=0.24$ (for H1) and $0.74$ (for H4), noting that we restrict our analysis to  times $t\in [150-200]$ (for H1) and $t\in [150-300]$ (H4). Comparisons of the time-resolved H1 and H4 raw and PINN data are provided in Supplementary Movies 1-4.

First, figure~\ref{fig:sturc_stat}(a,b) shows an instantaneous spanshot vertical velocity $w(x,z)$ taken in the mid-plane $y=0$ and at time $t=283$ in H4. The large-scale HW structures of the experiment (panel a) are faithfully reconstructed by the PINN (panel b) with much less small-scale structure, which we identify as experimental noise, violating the physical constraints and increasing the loss in \eqref{eq:loss_func}. 
This instantaneous $w$ snapshot
closely matches the `confined Holmboe instability' mode predicted by ~\cite{lefauve2018structure} (see their figure~9m) from a stability analysis of the mean flow in the same dataset H4. Such `clean', noise-free augmented experimental data will improve the three-dimensional structure of HW in \S~\ref{sec:vortex}.

Second, figure~\ref{fig:sturc_stat}(c-f) illustrates how the PINN is able to correct the inevitable distortion of experimental data along the spanwise, scanning direction (recall \S~\ref{sec:method:exp} and figure~\ref{fig:sid}), also clearly visible in Supplementary Movie 4. 
The top view shows a snapshot of the horizontal plane $\rho(x,y)$ sampled at $z=0.1$ near the density interface of neutral density where $\langle \rho\rangle_{x,y,t} \approx 0$. The top panels (c,e) show two successive volumes taken at time $t=178.9$ (forward scan) and $181.2$ (backwards scan) in flow H1, where the distortion is greatest (due to $\Delta t=2.3$ twice as large as in H4), while the bottom panels (d,f) show the respective instantaneous volumes output by the PINN. The original data make the peaks and troughs of this right-travelling HW mode appear alternatively slanted to the right during a forward scan, and to the left during a backward scan (see slanted black arrows). This distortion is successfully corrected by the PINN, which shows HW propagating along $x$ with a phase plane normal to $x$ (straight arrows), as predicted by theory \citep{ducimetiere2021effects}.

\begin{figure}
    \centering
    \includegraphics[width=.36\linewidth, trim=5mm 0mm 3mm 0mm, clip]{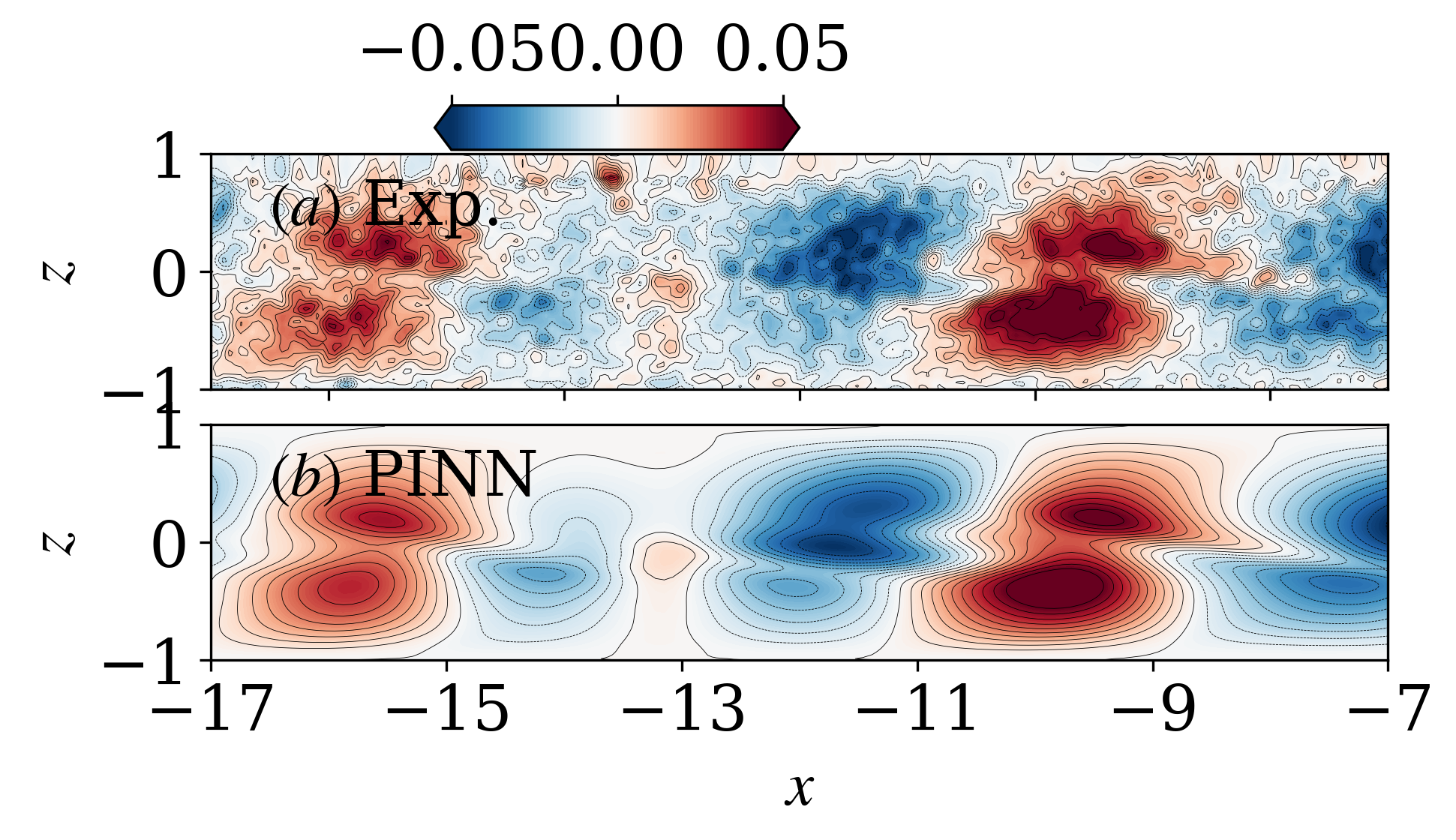}
    \includegraphics[width=.63\linewidth, trim=5mm 0mm 0mm 0mm, clip]{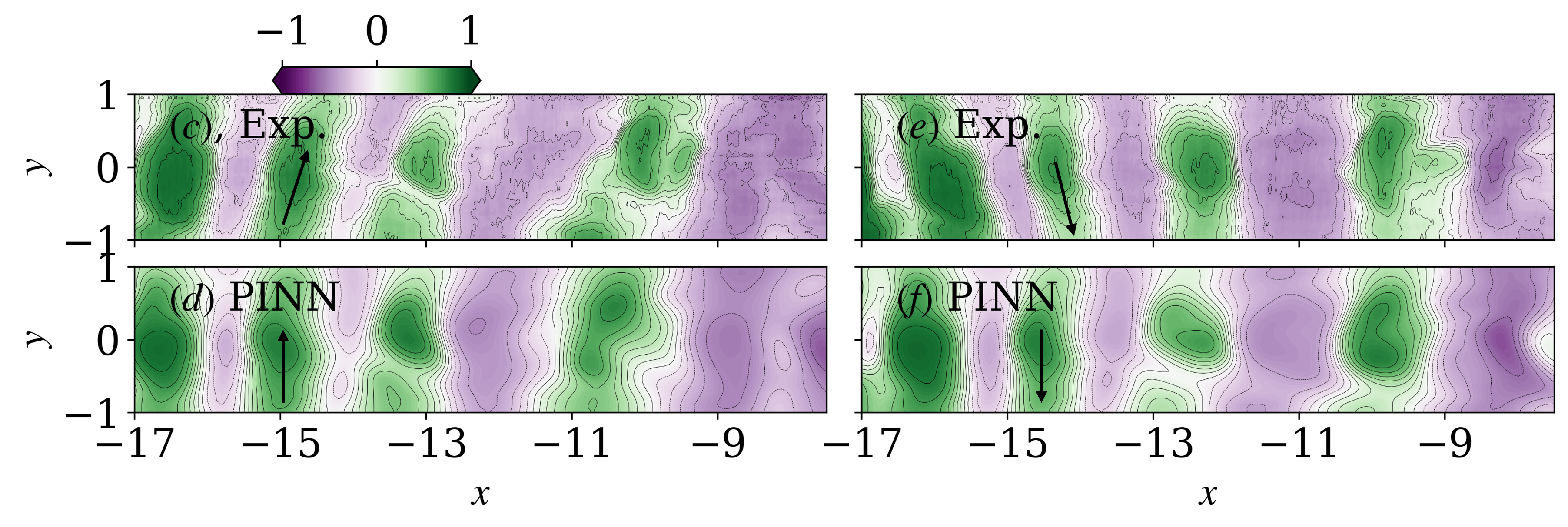}
    \caption{Improvement of instantaneous snapshots: (a,b) vertical velocity $w(x,y=0,z)$ in H4 at $t=283$, and (c-f) density just above the interface $\rho(x,y,z=0.1)$ in H1 at (c,d) $t=178.9$ (forward scan) and (e,f) $t=181.2$ (backward scan). }
    \label{fig:sturc_stat}
\end{figure}
\begin{figure}
    \centering
    \includegraphics[width=.95\linewidth, trim=0mm 0mm 0mm 0mm, clip]{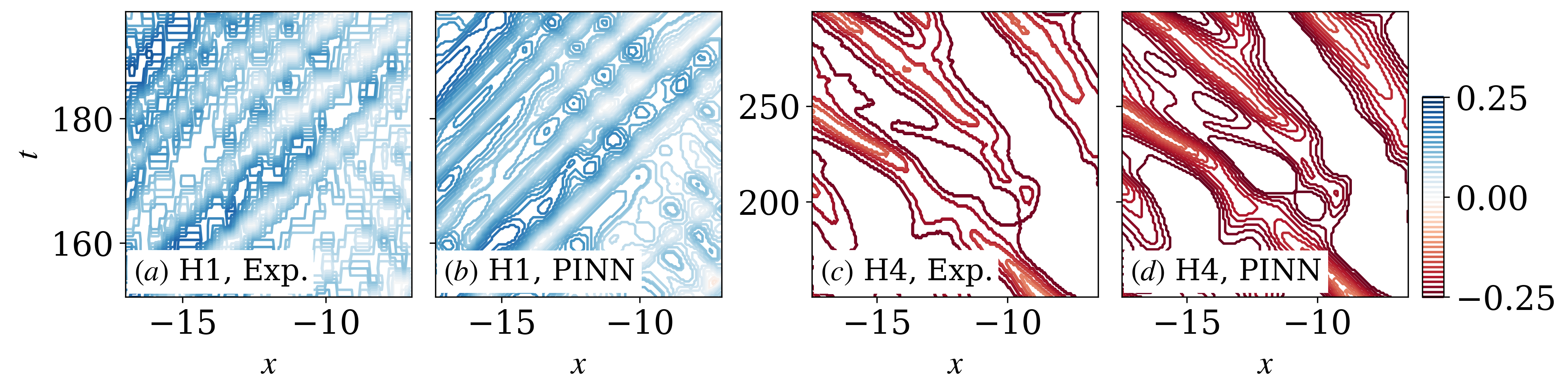}
    \caption{Improvement of the spatio-temporal diagrams of the interface height $\eta(x,t)$ for (a,b) H4 and  (c,d) H1, capturing the characteristics of HW propagation.\vspace{-0.2cm} }
    \label{fig:h0}
\end{figure}

Third, figure~\ref{fig:h0} compares the spatio-temporal  diagrams of interface height $\eta(x,t)$, defined as the vertical coordinate where $\rho(x,y=0,\eta,t)=0$. The characteristics showing the propagation of HWs in experiments (panels a,c) are largely consistent with the PINN results (panels b,d) but noteworthy differences exist. The determination of a density interface ($\rho= 0$) is particularly subject to noise due to a low signal-to-noise ratio when the signal approaches zero. Here we see that the contours are rendered much less jagged by the PINN. The 10-fold increased temporal resolution in H1 ($\Delta t=0.24$ vs 2.3) also allows a much clearer picture of the characteristics (panels c-d), which propagate in both directions, with sign of interference. By contrast,  H4 has a single leftward propagating mode as a result of the density interface ($\eta\approx-0.2$) being offset from the mid-point of the shear layer ($u=0$ at $z\approx -0.1$), as explained by \citep{lefauve2018structure}. However, the reason behind this offset was not elucidated by experimental data alone, and will be elucidated by the pressure field in \S~\ref{sec:pres}.

Fourth, we quantify and elucidate the magnitude of the PINN correction through the root-mean-square difference $d$ between the $\boldsymbol{u}$ and $\rho$ of the raw experiment and the reconstruction. We find $d=0.069$ for H1 and $d = 0.029$ for H4. We identify at least two sources contributing to $d$: (i) the noises in planar PIV/LIF measurements and (ii) the distortion of the volumes caused by scanning in $y$. As a baseline, we also compare a third, fully laminar dataset (L1) at $Re,\theta=(398, 2^\circ)$ \citep{lefauve2019regime} having a PINN correction of $d=0.030$. In L1, $d$ is primarily attributed to (i), since this simple steady, stable flow renders (ii) negligible. The energy spectrum of L1 in \cite{lefauve_experimental2_2022} (their figure 4a) highlighted the presence of small-scale measurement noise. We are confident that cause (i) remains relatively unchanged in H1 and H4, resulting in $d$ being predominantly explained by (i) in H4, while the larger spanwise distortion (ii) must be invoked to explain the larger $d$ in H1, consistent with an increasing interface variation and a scanning time that is twice  as slow as in figure~\ref{fig:sturc_stat}(c-f).

\subsection{Improved three-dimensional vortical structures}\label{sec:vortex}

Previous studies of this HW dataset in SID revealed how, under increasing turbulence levels quantified by the product $\theta\,Re $, the relatively weak three-dimensional Holmboe vortical structures evolve into pairs of counter-propagating turbulent hairpin vortices \citep{jiang_evolution_2022}. The particular morphology of these vortices conspires to entrain and stir fluid into the mixed interfacial region, elucidating a key mechanism for shear-driven mixing \citep{riley2022does}. However, \cite{jiang_evolution_2022} alludes to limitations in the signal-to-noise ratio to accurately resolve the structure of the weakest nascent Holmboe vortices, as shown, e.g., in a visualisation of the $Q$-criterion in their figure 1.

\begin{figure}
    \centering
    \includegraphics[width=0.99\linewidth, trim=0mm 0mm 0mm 0mm, clip]{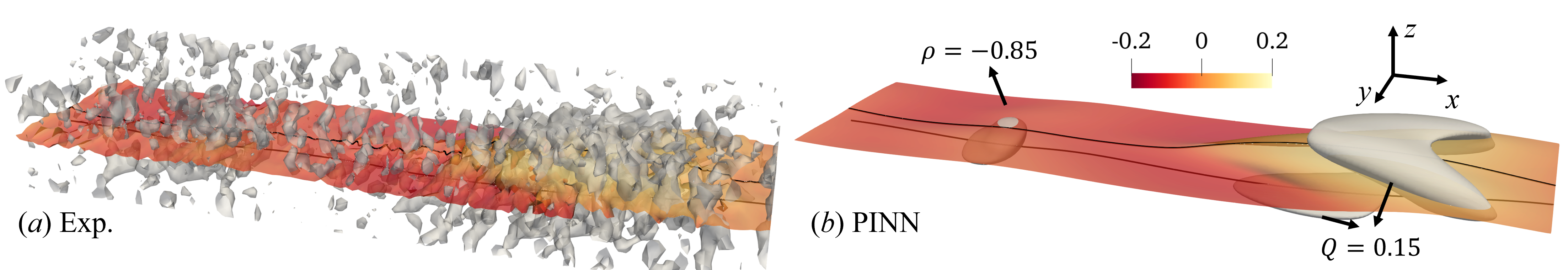}
    \caption{Three-dimensional vortex and isopycnal surfaces in H4 at $t=283$: (a) Exp. vs (b) PINN. 
    The grey iso-surfaces show $Q=0.15$, while the colours show the isopycnal $\rho=-0.85$ and its vertical position $z \in [-0.2,0.2]$.
    The two black lines are the isopcynals $\rho=-0.85$ and $\rho=0$ in the mid-plane $y=0$. \vspace{-0.2cm}}
    \label{fig:Q3D}
\end{figure}

Figure~\ref{fig:Q3D} shows how noise-free PINN data can uncover the vortex kinematics of this experimental HW in H4 by visualising an instantaneous $Q=0.15$ isosurface (in grey), where $Q=(1/2)(||\mathsfbi{W}||^2-||\mathsfbi{E}||^2)$, using the Frobenius tensor norm of the strain rate $\mathsfbi{E} =(1/2)(\boldsymbol\nabla\boldsymbol{u} + (\boldsymbol\nabla\boldsymbol{u})^T)$ and rotation rate  $\mathsfbi{W} =(1/2)(\boldsymbol\nabla\boldsymbol{u} - (\boldsymbol\nabla\boldsymbol{u})^T)$ \citep{hunt1988eddies}. A density isopycnal surface just above the density interface is superposed, with yellow to red shading denoting its vertical position. 

The unsmoothed experiment data (panel a) are highly fragmented rendering these weak individual vortices unrecognisable. A similar noise pattern in $Q$ is observed in the fully laminar dataset L1, not shown here,  confirming its unphysical nature.
The PINN data (panel b) effectively filters out the noise and shows well-organised vortices on either side of the isopycnal surface.

The interfacial vortices are flat and appear to be formed when  new wave crests appear (see left side of panel b), acting to lift and drop the interfaces, as the HW propagates (in this case, as a single left-going mode). Near the top of larger-amplitude isopycnal crests, a $\Lambda$-shape vortex is formed, which may eject wisps of relatively mixed fluid at the interface (low $|\rho|$) up into the unmixed region (high $|\rho|$), leading to the upward deflection of isopycnals. This locally `anti-diffusive' process is an example of scouring-type mixing typical of Holmboe waves \citep{salehipour_turbulent_2016,caulfield2021layering} and allows a density interface to remain sharp. This $\Lambda$-shape vortex causes the distance between the $\rho=-0.85$ and $0$ isopycnals (the two black lines) to increase, indicating enhanced mixing in this region.

These findings confirm the hypothesis of \cite{jiang_evolution_2022} and provide direct evidence for the existence of $\Lambda$-vortices in HW experiments.
Similar $\Lambda$-vortices have also been observed in more idealised numerical simulations of sheared turbulence ~\citep{watanabe_hairpin_2019}, proving their relevance beyond the SID geometry alone and the importance of correctly identifying these coherent structures in experimental data.

\subsection{Improved energy budgets and mixing efficiency}
\label{sec:eng_budget}

The energetics of turbulent mixing in stratified flows is a major topic of research in environmental fluid mechanics~\citep{colm2020open,dauxois_confronting_2021}. The energy budgets of datasets H1 and H4 were investigated in \cite{lefauve2019regime} and \cite{lefauve_experimental2_2022} but we will show that the PINN data can overcome limitations in spatial resolution (especially in $y$), in the relatively low signal-to-noise ratio of perturbation variables in HWs, and other limitations inherent to calculating energetics from experiments.%

\begin{figure}
    \centering
    \includegraphics[width=0.99\linewidth, trim=0mm 0mm 0mm 0mm, clip]{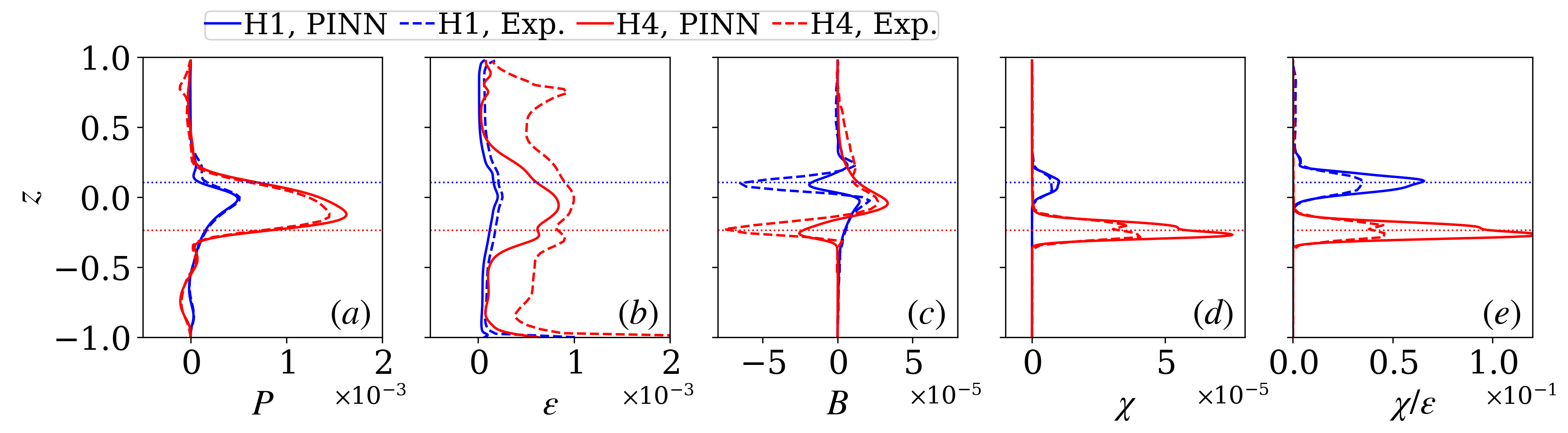}
    \caption{Improved energy budgets in H1 and H4: vertical profiles of (a) production $P$; (b) dissipation $\varepsilon$; (c) buoyancy flux $B$; (d)  scalar dissipation  $\chi$; (e) mixing efficiency $\chi/\varepsilon$, as defined in \eqref{eq:energ_budget}. The blue and red dotted line shows the mean density interface $\langle \rho \rangle=0$. \vspace{-0.2cm}}
    \label{fig:energ_bud}
\end{figure}

Figure~\ref{fig:energ_bud} shows four key terms in the budgets of turbulent kinetic energy (TKE) $K^\prime=(1/2) (\boldsymbol{u}'\cdot  \boldsymbol{u}')$ and turbulent scalar variance (TSV) $K^\prime_\rho = (\Ri/2) (\rho')^2$, namely the production $P$ and dissipation $\varepsilon$ of TKE, the buoyancy flux $B$ (exchanging energy between TKE and TSV) and the dissipation of TSV $\chi$, defined as in \cite{lefauve_experimental2_2022}:
\begin{equation}
\label{eq:energ_budget}
    P \equiv - \langle u'v'\p_y \bar{u}+ u'w' \p_z \bar{u}\rangle , \ \ 
    \varepsilon \equiv  \frac{2}{\Rey} \langle  ||\mathsfbi{E}'||^2 \rangle , \ \ 
    B  \equiv  \Ri \langle  w' \rho'\rangle , \   \
    \chi \equiv \frac{\Ri}{\Rey \, \Pran}\langle  |\boldsymbol{\nabla} \rho'|^2\rangle ,
\end{equation}
where fluctuations (prime variables) are computed around the $x,t$ averages (overbars), as in $\phi'=\phi-\bar{\phi}$, and $\langle \, \rangle$ denote averaging over $x,y,$ and $t$ (but not $z$).

The vertical profiles of TKE production $P(z)$ (panel a) show little difference between the PINN data (solid lines) and the experimental data (dashed lines), whether in H1 (blue) or H4 (red). We rationalise this by the fact that $P$ does not contain any derivatives of the velocity perturbations, and is thus less affected by small-scale noise. By contrast, the TKE dissipation $\varepsilon(z)$ (panel b) does contain gradients, and, consequently, greater differences between PINN and experiments are  found. In H4, the noise in experiments overestimates dissipation, especially away from the interface and near the walls ($|z|\gtrsim 0.5$) where turbulent fluctuations are not expected. 

The buoyancy flux $B(z)$ (panel c) and TSV dissipation $\chi(z)$ peak at the respective density interfaces of H1 ($z\approx 0.1$) and H4 ($z\approx -0.2$) and significant differences between experiments and PINN are again found. While experiments typically overestimate the magnitude of $B$ (which does not contain derivatives), they underestimate $\chi$, which contains derivatives of $\rho'$ that occur on notoriously small scales in such a high $\Pran = 700$ flow. These derivatives are expected to be better captured by the PINN as a consequence of its ability to super-resolve the density field. We also note the locally \textit{negative} values of $B$ at the respective interfaces, which confirm the scouring behaviour of HWs \citep{zhou2017diapycnal}. The mixing efficiency, defined as the ratio of TSV to TKE dissipation $(\chi/\varepsilon)(z)$ (panel e),  is about twice as high in the PINN reconstructed data than in the experiments, peaking sharply at $\chi/\varepsilon\approx 0.06$ in H1 and $\chi/\varepsilon\approx0.12$ in H4 at the respective density interfaces (note that volume-averaged values are an order of magnitude below). The ability of the PINN to correct mixing efficiency estimates is significant for further research into mixing of high$-\Pran$ turbulence. 

Finally,  $K^\prime$ and $K^\prime_\rho$ are expected to approach a statistical steady state when averaged over the entire volume $\mathcal{V}$ and over long time periods of order 100, such that the sum of all sources and sinks in their budgets should cancel \citep{lefauve_experimental2_2022}.
Importantly, the PINN predicts more plausible budgets than experiments, e.g. in H4, $\langle \partial_t K^\prime\rangle_{\mathcal{V},t}=-5.7\times 10^{-5}$ (eight times closer to zero than the experiment:  $-4.8\times 10^{-4}$) and $\langle \partial_t K^\prime_\rho\rangle_{\mathcal{V},t}=8.6\times 10^{-6}$ (five times closer to zero than the experiment  $-4.2\times 10^{-5}$).

\subsection{Revealing the latent pressure field}\label{sec:pres}

It is well known than an offset of the sharp density interface (here $\langle \rho \rangle= 0$) with respect to the mid-point of the shear layer (here $\langle u \rangle =0$) leads to asymmetric HWs \citep{lawrence_AHWS_1991}, where one of the travelling modes dominates at the expense of the other, which may disappear entirely, as in H4.  Recent direct numerical simulations (DNS) of SID revealed the non-trivial role of the pressure field in offsetting the density interface \citep{ZhuAtoufi2022} through a type of hydraulic jump appearing at relatively large duct tilt angle of $\theta=5^\circ$, as in H4. However, due to computational costs, these DNS were run at $\Pran=7$ (versus $700$ in experiments), and could thus not reproduce Holmboe waves. Here we demonstrate the physical insights afforded by the PINN reconstruction of $\Pran=700$ experimental data.

\begin{figure}
    \centering
    \includegraphics[width=.85\linewidth, trim=0mm 0mm 0mm 0mm, clip]{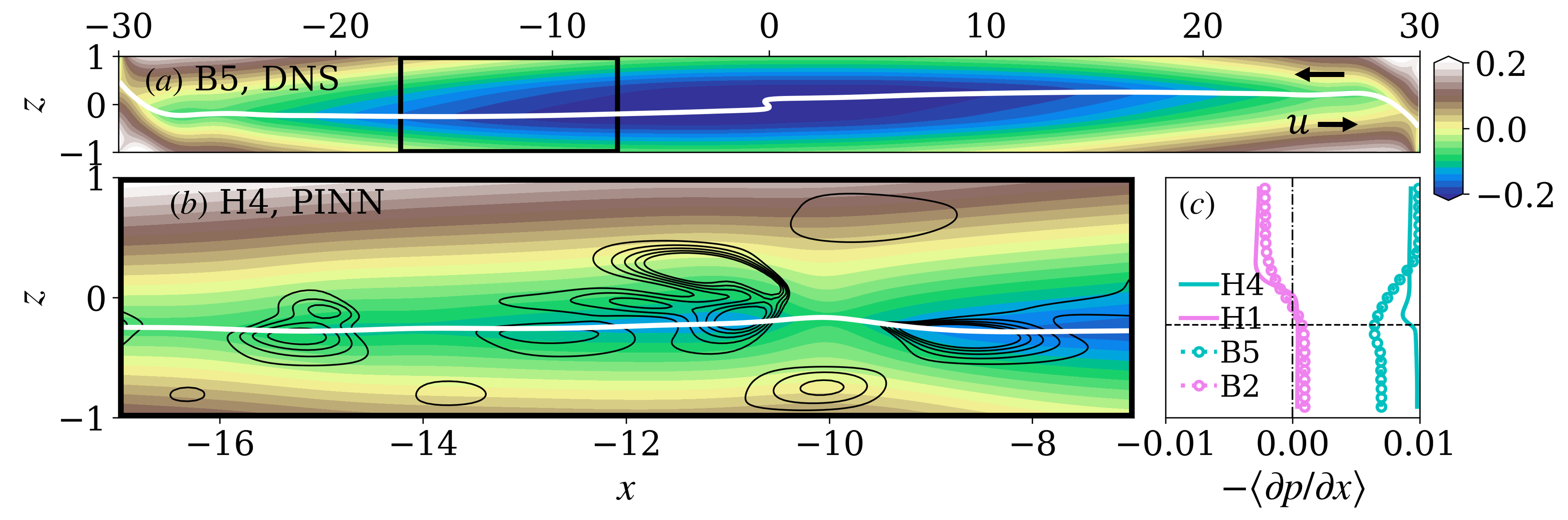}
    \caption{Prediction of the latent pressure: instantaneous pressure field in the mid-plane $y=0$ (colours). (a) DNS reproduced from case B5 of \citet{ZhuAtoufi2022} (whole duct shown). (b) H4 reconstructed by PINN showing the measured volume $x\in [-17,-7]$ only, indicated by a black box in (a), with $Q$-criterion black lines superimposed. The white solid lines indicate the density interface $\rho=0$. (c) Longitudinal pressure force $-\langle \partial_x p\rangle(z)$ in H1 and H4, compared with the closest respective DNS B2 and B5.  \vspace{-0.2cm}}
    \label{fig:pres}
\end{figure}

Figure~\ref{fig:pres}(a,b) compares the instantaneous non-dimensional pressure field predicted by the PINN in H4 (panel b) to that  DNS of \citet{ZhuAtoufi2022} (case B5, panel a) where the full duct geometry was simulated at identical $\theta= 5^\circ$ and slightly higher $\Rey= 650$ (vs 438).

Although the vastly different $\Pran$ led to slightly different flow states (HW in H4 versus a stationary wave in B5, leading to an internal hydraulic jump at $x=0$), the pressure distributions have clear similarities, as seen by comparing the black box of panel a to panel b. In both cases, a minimum pressure is found near the density interface, and a negative pressure (blue shades) is found nearer the centre of the duct $x>-10$. This minimum yields, in the top layer ($z>0$) on the left-hand side of the duct ($x<0$), a pressure that increases from right to left, i.e. in the direction of the flow, which slows down the upper layer over the second half of its transit along the duct. A symmetric situation occurs in the bottom layer on the right hand side of the duct. This behaviour was rationalised as the consequence of a hydraulic jump at $\theta =5^\circ$ \citep{atoufi2023stratified}, which was absent at $\theta =2^\circ$. 

This physical picture is confirmed by the pressure force $-\partial_x p(z)$ profiles in figure~\ref{fig:pres}(c). This panel shows that both H1 and B2 (at low $\theta$) have the typical favourable force $-\partial_x p(z)$ (positive in the lower layer, negative in the upper layer) expected of horizontal exchange flows. However, both H4 and B5 have an adverse pressure force in the upper layer on the left-hand side of the duct (i.e. $-\partial_x p>0$), which is typical downstream of a hydraulic jump, and results in the density interface being shifted down in this region, explaining the asymmetry of HWs.

The superposed $Q$-criterion lines to PINN in panel b also highlight that this low pressure zone is associated with intense vortices, which is consistent with the common vortex-pressure relation~\citep{hunt1988eddies}. %Jeong1995
We hypothesise that the lift-up of a three-dimensional HW in a shear layer causes the development of local high shears in proximity to the wave, which further evolve into $\Lambda$-vortices \citep{jiang_evolution_2022}.

\section{Conclusions}\label{sec:ccl}

In this paper, we applied a physics-informed neural network (PINN) that uses physical laws 
to augment experimental data and applied it to two stratified inclined duct datasets featuring symmetric and asymmetric Holmboe waves (HWs) at high Prandtl number $\Pran\approx700$. 

We first demonstrated in \S~\ref{sec:results_quality} the elimination of unphysical noise and of the spanwise distortion or wavefronts caused by the scanning data acquisition,  yielding cleaner, highly-resolved spatio-temporal wave propagation plots. This noise reduction allowed us in  \S~\ref{sec:vortex} to unambiguously detect weak but influential three-dimensional vortical structures, previously connected to higher-$\Rey$ turbulent structures, and to study their interaction with isopycnals. The accuracy of energy budgets was also improved in \S~\ref{sec:eng_budget} owing to the PINN noise removal and super-resolution capabilities, especially for terms involving the computation of small-scale derivatives such as the dissipation of turbulent kinetic energy and scalar variance. Mixing efficiency was revealed to be twice as high as suggested by raw experimental data, locally peaking at $0.06$ in the symmetric HW case and $0.12$ in the asymmetric HW case. Finally, additional physics were uncovered in \S~\ref{sec:pres} through the latent pressure field, confirming the existence of a pressure minimum towards the centre of the duct observed in simulation data. This minimum was linked to the existence of a hydraulic jump, offsetting the density interface, which is key to explain the presence or absence of asymmetric HW in our data.

These results mark a significant step forward in experimental fluid mechanics, and hold particular promise for the study of density-stratified turbulence and mixing from state-of-the-art laboratory data. 
Future work should attempt to reconstruct the more challenging turbulent flows at higher values of $\theta \, \Rey$ than done in this paper. This should resolve the large spectrum of turbulent scales below experimental resolution, which is especially relevant for scalar dissipation and mixing rates given the separation of scales at high $\Pran$.

\vspace{0.1cm}

We acknowledge the ERC Horizon 2020 Grant No 742480 `Stratified Turbulence And Mixing Processes'. A.L. acknowledges a NERC Independent Research Fellowship (NE/W008971/1). \vspace{0.1cm}

Declaration of interest: The authors report no conflict of interest.

\bibliographystyle{jfm}
\bibliography{References.bib}

\end{document}